\documentclass[prb,preprint,showpacs, floatfix,superscriptaddress]{revtex4-1}
\usepackage{graphicx}
\usepackage{dcolumn}
\usepackage{bm,amsmath,verbatim}

\def\be{\begin{equation}}
\def\ee{\end{equation}}
\def\bea{\begin{eqnarray}}
\def\eea{\end{eqnarray}}
\def\bse{\begin{subequations}}
\def\ese{\end{subequations}}

\def\be{\begin{eqnarray}}
\def\ee{\end{eqnarray}}

\begin{document}

\title{Topologically non-trivial superconductivity in spin-orbit coupled systems: Bulk phases and quantum phase transitions}

\author{Sumanta Tewari}
\affiliation{Department of Physics and Astronomy, Clemson University, Clemson, SC 29634}
\author{Tudor D. Stanescu}
\affiliation{Department of Physics, West Virginia University, Morgantown, WV 26506}

\author{Jay D. Sau}
\affiliation{Condensed Matter Theory Center and Joint Quantum Institute, Department of Physics, University of
Maryland, College Park, MD 20742}
\author{S. Das~Sarma}
\affiliation{Condensed Matter Theory Center and Joint Quantum Institute, Department of Physics, University of
Maryland, College Park, MD 20742}
\date{\today}

\begin{abstract}
Topologically non-trivial superconductivity has been predicted to
occur in superconductors with a sizable spin-orbit
coupling in the presence of an external Zeeman splitting. Two such systems have been proposed: (a)
$s$-wave superconductor pair potential is proximity induced on a semiconductor, and (b) pair potential
naturally arises from an intrinsic $s$-wave pairing interaction.
As is now well known, such systems in the form of a 2D film or 1D nano-wires in a wire-network can be used for topological quantum computation.
When the external Zeeman splitting $\Gamma$ crosses a critical value $\Gamma_c$, the system passes from a regular superconducting phase to a non-Abelian topological
superconducting phase.
  In both cases (a) and (b) we consider in this paper the pair potential $\Delta$ is strictly $s$-wave in both the ordinary and the topological superconducting phases, which are separated by a topological quantum critical point at $\Gamma_c = \sqrt{\Delta^2 + \mu^2}$, where $\mu (>> \Delta)$ is the chemical potential.
  On the other hand, since $\Gamma_c >> \Delta$, the Zeeman splitting required for the topological phase ($\Gamma > \Gamma_c$) far exceeds the value ($\Gamma \sim \Delta$) above which an $s$-wave pair potential is expected to vanish (and the system to become non-superconducting) in the absence of spin-orbit coupling. We are thus led to a situation that the topological superconducting phase appears to set
in a parameter regime at which the system actually is non-superconducting in the absence of spin-orbit coupling. In this paper we
address the question of how a pure $s$-wave pair potential can survive a strong Zeeman field to give rise to a topological superconducting phase. We show that the spin-orbit coupling is the crucial parameter for the quantum transition into and the robustness of the
topologically non-trivial superconducting phase realized for $\Gamma >> \Delta$.
\end{abstract}

\maketitle

\section{Introduction}

Recently topologically non-trivial superconductivity has been theoretically predicted to occur in two classes of systems with spin-orbit (SO) coupling. They are: (a) SO coupled semiconductors in which $s$-wave superconducting pair potential is induced by the proximity effect \cite{sau1, Ann, Long-PRB} and (b) SO coupled systems with superconductivity due to intrinsic $s$-wave pairing interaction \cite{Parag, Sato-Fujimoto}. A third system - surface of a 3D strong topological insulator (TI) - can also support topological superconductivity when the latter is proximity induced \cite{fu_prl'08}. In this paper we will
ignore the latter and concentrate only on (a) and (b). In both classes (a) and (b) the SO coupling is a consequence of the
breakdown of the structural space inversion (SI) symmetry. We will take the resultant SO coupling to be of the Rashba type. A system in class (a) can be artificially grown as a heterostructure consisting of a SO coupled semiconductor in proximity contact with a $s$-wave superconductor \cite{Long-PRB}. An example of class (b) is a non-centrosymmetric superconductor \cite{Frigeri} or an $s$-wave Feshbach resonant system with an accompanying spin-orbit coupling and Zeeman splitting both of which can be created in a cold fermion atomic system \cite{Chuanwei1,Chuanwei2}.
There is also another class (class (c)) of topological superconductors with non-Abelian statistics, which have been studied extensively in the recent literature. Here the topological nature arises entirely from intrinsic chiral $p$-wave superconductivity without having any underlying $s$-wave superconductivity in the system.  Some examples of this class (c) are the even-denominator 5/2 fractional quantum Hall effect \cite{DasSarma-2005}, superconducting strontium ruthenate \cite{DasSarma-2006}, A-phase of superfluid He-3 \cite{Volovik}, and superfluid ultracold  fermionic gases based directly on the $p$-wave  Feshbach resonance \cite{Tewari-2007}.  These class (c) non-Abelain superconductors are effectively equivalent to being spinless, i.e. completely spin-polarized, and are therefore immune to any Zeeman splitting to the leading order.  We do not discuss the class (c) systems in this paper since the main conceptual issue being addressed in this paper does not apply to these systems.

\begin{figure}[tbp]
\begin{center}
\includegraphics[width=0.65\textwidth]{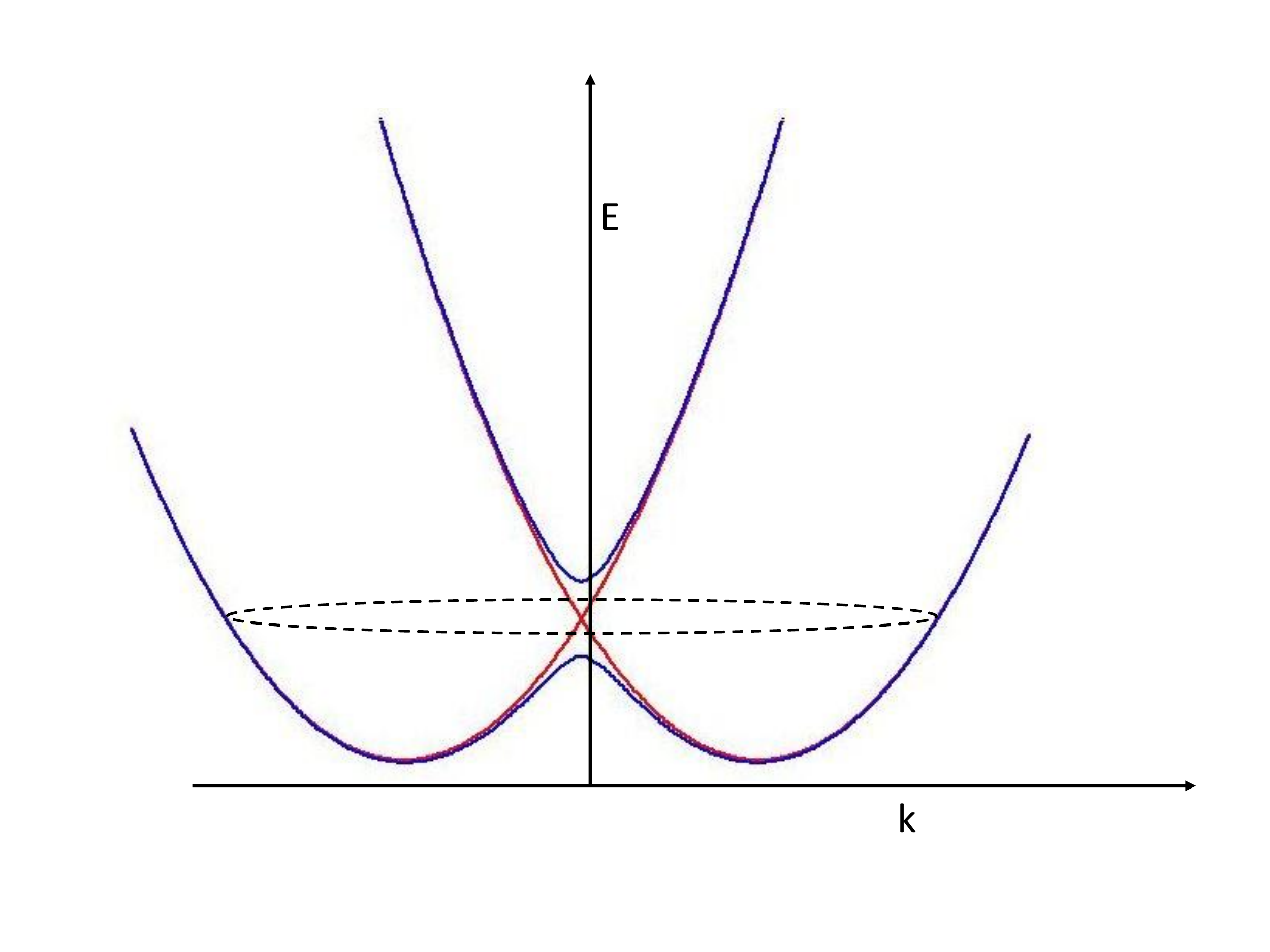}
\end{center}
\caption{(Color online) The two spin-orbit bands with (blue curves) and without (red curves) Zeeman splitting are shown schematically. With
Zeeman splitting the bands have a band gap at the origin. When the Zeeman splitting is large enough so that the chemical potential (dotted circle) lies in the gap, the system has only one Fermi surface and the ordinary superconducting phase gives way to a topologically non-trivial superconducting phase.}
\label{Fig0}
\end{figure}

The topologically non-trivial superconducting systems mentioned above are
characterized by order parameter defects, such as a vortex and a sample edge, which carry
a unique bound state at zero excitation energy \cite{Read}. These bound zero energy modes, called Majorana fermion modes after E. Majorana \cite{Majorana}, can actually be thought of as particles which are their own anti-particles \cite{Wilczek-3}. In other words, they are represented by second quantized operators $\gamma$ which satisfy the hermiticity condition $\gamma^{\dagger}=\gamma$. This is strikingly different from the regular fermionic modes for which the second quantized operators $c^{\dagger} \neq c$. Because of the existence of the zero-energy Majorana modes, a 2D non-Abelian topological superconductor with $2N$ vortices each carrying a single Majorana mode constitutes a system with a $2^N$-fold ground state degeneracy protected by an excitation gap \cite{Ivanov}. When a Majorana fermion mode in a given order parameter defect is adiabatically moved (braided) around another Majorana mode, the initial state transforms
into another one which is a different linear combination in the same ground state manifold. These unitary transformations within the ground state manifold are manifestations of the non-Abelian statistics of the Majorana fermions. The non-Abelian statistics of the Majorana fermions has recently come under intense focus because of its potential application in fault tolerant topological quantum computation (TQC) \cite{Kitaev, nayak_RevModPhys'08}. Both 2D spin-orbit coupled superconducting films and its 1D version as quantum nanowires in a wire-network have been proposed as potential platforms for TQC \cite{sau1, Roman, Oreg, Alicea-Network, Sau-Interferometry, Sau-UQC, Beenakker-UQC,Roman-Tudor}. In this paper we will not discuss the Majorana fermion modes or their potential application to TQC. Instead we will focus on the bulk superconducting phases and the quantum phase transitions (QPT) by exploring the interplay of the spin-orbit coupling, Zeeman splitting, and the $s$-wave pairing interactions.


 Both systems in classes (a) and (b) are in an ordinary (non-topological) superconducting phase in the absence of an externally imposed Zeeman splitting. The Zeeman splitting, which can be applied either by a parallel magnetic field \cite{Alicea-Tunable} or by the proximity effect of a nearby magnetic insulator \cite{sau1,Ann,Long-PRB}, creates a gap between the two spin-orbit bands as shown in Fig.~\ref{Fig0}. From mean field calculations \cite{sau1,Long-PRB} it is clear that when such a Zeeman splitting $\Gamma$ crosses a critical value, $\Gamma_c = \sqrt{\Delta^2 + \mu^2}$, where $\Delta$ is the $s$-wave superconducting pair potential and $\mu$ is the chemical potential, the system makes a transition to a topological non-Abelian superconducting phase. The critical Zeeman splitting corresponds to the value at which the underlying Fermi surface in the absence of superconductivity shifts from being at both SO bands to occupying only the lower SO band (Fig.~\ref{Fig0}). Thus, at this value of the Zeeman splitting, the underlying system changes from having two Fermi surfaces (one in each band) to having just one.

 Since $\mu$ usually far exceeds $\Delta$, the critical value of the Zeeman splitting $\Gamma_c$ also far exceeds $\Delta$. Therefore, in the absence of SO coupling, $\Gamma_c$ far exceeds the Zeeman splitting ($\Gamma \sim \Delta$) above which an $s$-wave pair potential should decay to zero. The loss of an $s$-wave pair potential due to a strong Zeeman splitting is due to the fact that for $\Gamma \gtrsim \Delta$ forming a spin-singlet pair potential with zero net momentum is impossible (for the discussion of topological phase transition induced by Zeeman splitting we will ignore the states with non-zero values of the Cooper pair momentum). Now let us emphasize the fact that in
 both cases we consider in this paper (superconducting pair potential proximity induced, and superconducting pair potential due to
 an intrinsic on-site pairing interaction), the pair potential is strictly $s$-wave in both the ordinary and the topologically non-trivial
  superconducting phases. The fact that the pair potential is $s$-wave
 when it is proximity induced from a $s$-wave superconductor is self-evident. That it remains $s$-wave (and is not a mixture of $s$- and $p$-wave due
 to the SO coupling) even when the pair potential is due to an intrinsic on-site pairing interaction is not so obvious. In this case, the pure $s$-wave symmetry
 of the pair potential follows from the fact that the
 intrinsic pairing interaction we consider is spatially local, and thus forming a $p$-wave component of the pair potential is forbidden by the fermion anticommutation relation (for a more detailed discussion see Sec. V).
  Since the pair potential is purely $s$-wave in both classes (a) and (b) and in both phases (ordinary and topological) in each, how an $s$-wave pair potential survives a strong Zeeman splitting $\Gamma >> \Delta$ to realize the topologically non-trivial phase is the central conceptual question we address in this paper.

 The basic conceptual issue being discussed here is the topic often alluded to as the Chandrasekhar-Clogston (CC) limit \cite{Chandrasekhar,Clogston} in ordinary $s$-wave superconductivity, which states, in effect, that an $s$-wave superconductor, where the Cooper pairing is between spin-up and spin-down electrons near the Fermi surface, cannot withstand a Zeeman splitting larger than the superconducting gap. This is because then spin splitting exceeds the superconducting gap energy, making it impossible for a superconducting ground state to develop.  On first sight, it appears that the condition on the Zeeman splitting needed for superconductivity in Refs. [1-4] far exceeds this limit, thus destroying all superconductivity!  This has caused some confusion about the very existence of the topological superconducting phase using
 either a heterostructure \cite{sau1,Ann,Long-PRB} where $s$-wave superconductivity is induced by proximity effect or using SO-coupled systems with intrinsic $s$-wave pairing interactions \cite{Parag}.

 The mean field calculations of Ref.~[1-4] are not enough to resolve this question. This is because a mean field theory is not just
a postulate to assume the existence of a mean field pair potential in the Hamiltonian H as is done in these works; one is also required to establish
 the finiteness of the pair potential by satisfying the self-consistent gap equation.
In other words, we need to satisfy the Bardeen-Cooper-Schrieffer (BCS) self-consistent gap equation with a strong
Zeeman potential ($\Gamma > \sqrt{\mu^2 +\Delta^2}$) to check if a non-zero $s$-wave pair potential $\Delta$ gives a consistent solution. This will ensure that the mean field H in Refs. [1-4] is not flawed to begin with,
and our BdG solution of the Majorana fermion is not
a spurious mathematical result with no physical connection.
In this paper we perform this study by self-consistently solving the appropriate BCS gap equations in the presence of attractive $s$-wave pairing interaction, SO coupling, and an externally applied Zeeman splitting. \emph{Note that satisfying such a gap equation is a requirement for superconducting pair potential when it is derived from the microscopic pairing interactions. However, when the pair potential is proximity induced on a spin-orbit coupled system by a nearby $s$-wave superconductor, the gap equation need not be satisfied. In this case, the SO coupled system simply `inherits' the pair potential of the nearby superconductor.}

 For the case of intrinsic pairing interactions, we show that the $s$-wave pair potential indeed remains non-zero even beyond the Zeeman splitting above which it would be lost in the absence of SO coupling. In fact, in the non-Abelian phase the non-zero value of the pair potential crucially depends on and \emph{ increases} with the SO coupling constant, which is consistent with the fact that it is zero in the absence of the SO coupling. A simple intuitive way to understand this starts by recalling how $s$-wave superconductivity is destroyed by a Zeeman splitting. In the absence of SO coupling, the two spin bands are shifted by an energy gap proportional to the Zeeman splitting $\Gamma$. With increasing $\Gamma$ it becomes increasingly difficult for the system to create $s$-wave spin-singlet pairs with zero net momentum. Finally, when $\Gamma$ crosses a value $\sim \Delta$ the $s$-wave pair potential vanishes. The critical Zeeman splitting, $\Gamma_c = \sqrt{\Delta^2 + \mu^2}$, needed for the topological phase transition is thus squarely beyond the acceptable Zeeman splitting the pair potential can sustain. It is now important to realize that, in the presence of the SO coupling, the two spin-orbit bands cannot simply be viewed as `spin-up' and `spin-down' bands. Instead, they both have a non-zero minority spin amplitude coexisting with the majority spin component. Therefore, even when the Zeeman splitting is large enough to make the Fermi surface lie only in the lower band,  spin-singlet $s$-wave pairs cannot be completely lost. If the superconductivity is due to an intrinsic pairing interaction the gap equation shows that the pair potential, though always non-zero, decays with increasing Zeeman couplings. However, the magnitude of the pair potential in the non-Abelian phase can be increased by increasing the magnitude of the spin-orbit coupling, which, therefore, enables a stable non-Abelian phase in the phase diagram .  On the other hand, if the superconductivity is due to the proximity effect, there is no requirement of satisfying the self-consistent gap equation. In this case, which applies to the heterostructure geometry,
the superconducting pair potential is simply `inherited' from the adjacent $s$-wave superconductor (Sec. VIII).

 \section{Hamiltonian} We assume that the quasi-2D electron system is described by the model Hamiltonian
\begin{equation}
H = H_0 + H_{SO} + H_{\Gamma} + H_{\rm int},   \label{Htot}
\end{equation}
where $H_0$ describes the bulk conduction electrons,
$H_{SO}$ is the spin-orbit interaction term, $H_{\Gamma}$ represents the Zeeman coupling,
and $H_{\rm int}$ represents the electron-electron interaction. Explicitly, we have
\begin{eqnarray}
H_0 &=& \sum_{{\bm p}, \sigma} \xi_{\bm p} c_{{\bm p} \sigma}^\dagger c_{{\bm p} \sigma},   \label{H0} \\
H_{SO} &=& \alpha \sum_{{\bm p}, \sigma, \sigma^\prime} c_{{\bm p}  \sigma}^\dagger \left(p_y\tau_x - p_x\tau_y\right)_{\sigma,\sigma^\prime} c_{{\bm p} \sigma^\prime},               \label{HSO}  \\
H_{\Gamma}&=&  \Gamma \sum_{\bm p} \left(c_{{\bm p} \uparrow}^\dagger c_{{\bm p} \uparrow} - c_{{\bm p} \downarrow}^\dagger c_{{\bm p} \downarrow}\right),      \label{HGamma} \\
H_{\rm int} &=& \frac{1}{2} \sum_{{\bm p}, {\bm p}^\prime, {\bm q}}\sum_{\sigma, \sigma^\prime} V(q) c_{{\bm p}+{\bm q}  \sigma}^\dagger c_{{\bm p}^\prime-{\bm q}  \sigma^\prime}^\dagger c_{{\bm p}^\prime \sigma^\prime} c_{{\bm p}, \sigma},   \label{Hint}
\end{eqnarray}
where $\xi_{\bm p}=p^2/2m -\mu$ is the bulk spectrum (measured relative to the chemical potential $\mu$), $\alpha$ is the strength of the Rashba spin-orbit coupling, $\Gamma$ represents the Zeeman field, $V(q)$ is the short-ranged interaction potential (we will later restrict ourselves only to an on-site pairing interaction, $V(q)$ independent of $q$, natural for $s$-wave order), $\tau_{x(y)}$ are Pauli matrices, and $c_{{\bm p} \sigma}^\dagger$ ($c_{{\bm p} \sigma}$) is the creation (annihilation) operator corresponding to the single-particle state with momentum ${\bm p}$ and spin $\sigma$. It is convenient to work in the spinor basis provided by the eigenfunctions $\phi_{\lambda}({\bm p})$ of the single-particle Hamiltonian $H_0+H_{SO}$,
\begin{equation}
\phi_{\lambda}({\bm p}) = \frac{1}{\sqrt{2}}\left(
\begin{array}{c}
1 \\
-i\lambda e^{i\theta_{\bm p}}
\end{array}\right),   \label{phi}
\end{equation}
where $\lambda = \pm$ and $e^{i\theta_{\bm p}} = (p_x+i p_y)/p$.  The corresponding eigenvalues are
\begin{equation}
\epsilon_{\lambda}({\bm p}) = \xi_{\bm p} + \lambda \alpha p. \label{epsilon}
\end{equation}
The electron $c$-operators can be expressed in terms of the annihilation operators $a_{{\bm p} \lambda}$ associated with the spinor eigenstates as
\begin{equation}
c_{{\bm p}\sigma} = \sum_{\lambda}\phi_{\lambda}({\bm p}, \sigma) a_{{\bm p} \lambda}.    \label{cphia}
\end{equation}
Using the spinor representation, the Hamiltonian becomes
\begin{eqnarray}
H_0&+&H_{SO} = \sum_{{\bm p},\lambda} \epsilon _{\lambda}({\bm p}) a_{{\bm p} \lambda}^\dagger a_{{\bm p} \lambda},   \label{H0lbd} \\
H_{\Gamma}&=& \Gamma \sum_{{\bm p},\lambda} a_{{\bm p}\lambda}^{\dagger} a_{{\bm p} -\lambda}  \label{HGammalbd} \\
H_{\rm int} &=& \frac{1}{2}\sum_{{\bm p}, {\bm p}^\prime, {\bm q}}\sum_{\lambda, \lambda^\prime, \mu, \mu^\prime} V(q) \chi_{\lambda \lambda^\prime}({\bm p}+{\bm q}, {\bm p})  \chi_{\mu\mu^\prime}({\bm p}^\prime-{\bm q}, {\bm p}^\prime) a_{{\bm p}+{\bm q} \lambda}^\dagger  a_{{\bm p}^\prime-{\bm q} \mu}^\dagger  a_{{\bm p}^\prime \mu^\prime} a_{{\bm p} \lambda},  \label{Hintlbd}
\end{eqnarray}
where
\begin{equation}
\chi_{\lambda\lambda^\prime} ({\bm p}_1,{\bm p}_2) = \frac{1}{2} \left(1 + \lambda\lambda^\prime e^{-i(\theta_{{\bm p}_1} -\theta_{{\bm p}_2})}\right)  \label{chi}
\end{equation}
is the scalar product of two spinors $\phi_{\lambda}({\bm p}_1)$ and $\phi_{\lambda^\prime}({\bm p}_2)$.

\section{Superconducting gap equations}

To derive the gap equations, we first introduce the regular and anomalous Green functions as
\begin{eqnarray}
{\cal G}_{\lambda\lambda^\prime}({\bm p}, \tau-\tau^\prime) &=& - \langle T_{\tau} a_{{\bm p} \lambda}(\tau) a_{{\bm p} \lambda^\prime}^\dagger(\tau^\prime)\rangle,   \label{Glbd} \\
{\cal F}_{\lambda\lambda^\prime}({\bm p}, \tau-\tau^\prime) &=& \lambda\langle T_{\tau} a_{{\bm p} \lambda}(\tau) a_{-{\bm p} \lambda^\prime}(\tau^\prime)\rangle,   \label{Flbd} \\
{\cal F}_{\lambda\lambda^\prime}^{+}({\bm p}, \tau-\tau^\prime) &=& \lambda\langle T_{\tau} a_{{\bm p} \lambda}^\dagger(\tau) a_{-{\bm p} \lambda^\prime}^\dagger(\tau^\prime)\rangle,   \label{Fxlbd}
\end{eqnarray}
where $T_{\tau}$ is the time ordering operator and the operators $a_{{\bm p} \lambda}(\tau)$ are in the Heisenberg representation. The correlation functions ${\cal F}_{\lambda\lambda^\prime}({\bm p}) = {\cal F}_{\lambda\lambda^\prime}({\bm p}, 0+)$ have the properties
\begin{eqnarray}
{\cal F}_{\lambda\lambda}(-{\bm p}) &=& -{\cal F}_{\lambda\lambda}({\bm p}), ~~~~~~~~~~~~~~~~~~~~~~ {\cal F}_{\lambda\lambda}^{+}(-{\bm p}) = {\cal F}_{\lambda\lambda}^*({\bm p}),  \label{Fminusp} \\
{\cal F}_{+-}(-{\bm p}) &=& {\cal F}_{-+}({\bm p}), ~~~~~~~~~~~~~~~~~~~~~~~ {\cal F}_{+-}^{+}(-{\bm p}) = {\cal F}_{-+}^*({\bm p}).  \label{Fminusp1}
\end{eqnarray}
The definitions (\ref{Flbd}) and (\ref{Fxlbd}) of the anomalous correlation functions follow the convention used by Gor'kov and Rashba \cite{Gorkov}.

Following the standard procedure, we write the equations of motion for the Green functions using the time evolution of the $a$-operators, $\partial_\tau a_{{\bm p}\lambda}(\tau) = [H, a_{{\bm p}\lambda}]$. We have
\begin{eqnarray}
&~&[-\partial_{\tau} - \epsilon_{\lambda}({\bm p})]{\cal G}_{\lambda\lambda^\prime}({\bm p}, \tau-\tau^\prime) - \Gamma {\cal G}_{-\lambda\lambda^\prime}({\bm p}, \tau-\tau^\prime) \nonumber \\
&~&~~~~~~~+ \sum_{{\bm q}, \mu, \mu_1, \lambda_1} V({\bm p}-{\bm q}) \chi_{\lambda\mu}({\bm p},{\bm q}) \chi_{\lambda_1\mu_1}(-{\bm p},-{\bm q}) \lambda_1\mu_1 {\cal F}_{\mu_1\mu}(-{\bm q}, 0+) {\cal F}_{\lambda_1\lambda^\prime}(-{\bm p}, \tau-\tau^\prime) = \delta_{\lambda\lambda^\prime} \delta(\tau-\tau^\prime), \nonumber \\
&~&[-\partial_{\tau} + \epsilon_{\lambda}({\bm p})]{\cal F}_{\lambda\lambda^\prime}^{+}(-{\bm p}, \tau-\tau^\prime) -\Gamma{\cal F}_{-\lambda\lambda^\prime}^{+}(-{\bm p}, \tau-\tau^\prime)    \label{EqMotion1} \\
&~&~~~~~~~ -\lambda \sum_{{\bm q}, \mu, \mu_1,\lambda_1} V({\bm p}-{\bm q}) \chi_{\mu\lambda}(-{\bm q},-{\bm p}) \chi_{\mu_1\lambda_1}({\bm q}, {\bm p}) \mu {\cal F}_{\mu\mu_1}(-{\bm q}, 0+)^+ {\cal G}_{\lambda_1\lambda}(-{\bm p}, \tau-\tau^\prime) = 0. \nonumber
\end{eqnarray}
The gap function can be defined as,
\begin{equation}
\Delta_{\lambda\lambda_1}({\bm p}) = \lambda_1 \sum_{{\bm q}, \mu, \mu_1} V({\bm p}-{\bm q}) \chi_{\lambda\mu}({\bm p},{\bm q}) \chi_{\lambda_1\mu_1}(-{\bm p},-{\bm q}) \mu_1 {\cal F}_{\mu_1\mu}(-{\bm q}, 0+).             \label{Delta}
\end{equation}
Introducing the definition of the gap function in Eq. (\ref{EqMotion1}), we have
\begin{eqnarray}
&~&[-\partial_{\tau} - \epsilon_{\lambda}({\bm p})]{\cal G}_{\lambda\lambda^\prime}({\bm p}, \tau-\tau^\prime) - \Gamma {\cal G}_{-\lambda\lambda^\prime}({\bm p}, \tau-\tau^\prime) + \sum_{\lambda_1}\Delta_{\lambda\lambda_1}({\bm p}) {\cal F}_{\lambda_1\lambda^\prime}^+(-{\bm p}, \tau-\tau^\prime) = \delta_{\lambda\lambda^\prime} \delta(\tau-\tau^\prime), \nonumber \\
&~&[-\partial_{\tau} + \epsilon_{\lambda}({\bm p})]{\cal F}_{\lambda\lambda^\prime}^{+}(-{\bm p}, \tau-\tau^\prime)-\Gamma {\cal F}_{-\lambda\lambda^\prime}^{+}(-{\bm p}, \tau-\tau^\prime) + \sum_{\lambda_1} \Delta_{\lambda_1\lambda}^* {\cal G}_{\lambda_1\lambda}({\bm p}, \tau-\tau^\prime) = 0. \label{EqMotion2}
\end{eqnarray}
Defining the Fourier transforms of the correlation functions in the usual way, ${\cal G}_{\lambda\lambda^\prime}({\bm p}, \tau) = k_B T \sum_n e^{-i \omega_n \tau} {\cal G}_{\lambda\lambda^\prime}({\bm p}, i\omega_n)$, the set of equations of motion can be expressed in a matrix form as
\begin{equation}
\left(
\begin{array}{cccc}
i\omega_n-\epsilon_+ & -\Gamma & \Delta_{++} & \Delta_{+-} \\
-\Gamma &  i\omega_n-\epsilon_- &   \Delta_{-+} & \Delta_{--} \\
\Delta_{++}^* & \Delta_{-+}^* &      i\omega_n+\epsilon_+ & -\Gamma \\
\Delta_{+-}^* & \Delta_{--}^* & -\Gamma &   i\omega_n+\epsilon_-
\end{array}
\right)
\left(
\begin{array}{c}
{\cal G}_{++} \\
{\cal G}_{-+} \\
{\cal F}^{+}_{++} \\
{\cal F}^{+}_{-+}
\end{array}
\right) =
\left(
\begin{array}{c}
1 \\
0 \\
0 \\
0
\end{array}
\right),     \label{MEqns}
\end{equation}
where the arguments of the Green functions have been omitted for simplicity. A similar set of equations, which can be obtained from (\ref{MEqns}) by switching the $+$ and $-$ labels,  couples ${\cal G}_{--}$, ${\cal G}_{+-}$, ${\cal F}_{--}$, and ${\cal F}_{+-}$. The superconducting spectrum can be obtained from the condition that the determinant of the $4\times 4$ matrix in Eq.  (\ref{MEqns}) vanish taking  $i\omega_n \rightarrow E$. Also, by solving the system of equations of motion for ${\cal F}_{\lambda\lambda^\prime}$ and introducing the solutions in Eq. (\ref{Delta}) we obtain the self-consistent gap equations. In general, we have $\Delta_{\lambda\lambda}({\bm p}) = i e^{-i\theta_{\bm p}} [\Delta_{0s}({\bm p}) +\lambda \Delta_{0a}({\bm p})]$, where the symmetric and antisymmetric components of the diagonal gap functions are
\begin{eqnarray}
\Delta_{0s}({\bm p}) &=& -\sum_{\bm q} \frac{V({\bm p}-{\bm q})+V({\bm p}+{\bm q}) }{2} \frac{i}{2}[{\cal F}_{++}(-{\bm q},0) +{\cal F}_{--}(-{\bm q},0) ] e^{i\theta_{\bm q}},   \label{del0s} \\
\Delta_{0a}({\bm p}) &=& -i \sum_{\bm q} \frac{V({\bm p}-{\bm q})-V({\bm p}+{\bm q}) }{2}  \left[({\cal F}_{++}-{\cal F}_{--})\cos(\theta_{\bm p}-\theta_{\bm q})e^{i\theta_{\bm q}}  + i({\cal F}_{+-}-{\cal F}_{-+})\sin(\theta_{\bm p}-\theta_{\bm q})e^{i\theta_{\bm q}} \right].\nonumber\\  \label{del0a}
\end{eqnarray}
Similarly, the off-diagonal gap functions can be expressed as  $\Delta_{\lambda~ -\lambda}({\bm p}) = i e^{-i\theta_{\bm p}} [\Delta_{1s}({\bm p}) +\lambda \Delta_{1a}({\bm p})]$, with
\begin{eqnarray}
\Delta_{1s}({\bm p}) &=& -\sum_{\bm q} \frac{V({\bm p}-{\bm q})-V({\bm p}+{\bm q}) }{2} \frac{i}{2}[{\cal F}_{+-}(-{\bm q},0) +{\cal F}_{-+}(-{\bm q},0) ] e^{i\theta_{\bm q}},   \label{del1s} \\
\Delta_{1a}({\bm p}) &=& -i \sum_{\bm q} \frac{V({\bm p}-{\bm q})-V({\bm p}+{\bm q}) }{2}  \left[i ({\cal F}_{++}-{\cal F}_{--})\sin(\theta_{\bm p}-\theta_{\bm q})e^{i\theta_{\bm q}}  + ({\cal F}_{+-}-{\cal F}_{-+})\cos(\theta_{\bm p}-\theta_{\bm q})e^{i\theta_{\bm q}} \right].\nonumber\\  \label{del1a}
\end{eqnarray}
Note that $\Delta_{js}(-{\bm p})= \Delta_{js}({\bm p})$ and  $\Delta_{ja}(-{\bm p})= -\Delta_{ja}({\bm p})$, i.e., $\Delta_{js}$ and $\Delta_{ja}$ represent the singlet and triplet components of the gap functions, respectively.

\section{Assumption of local interaction}
Instead of solving the
 complicated coupled set of gap equations above,
 we simplify matters by considering the case of strictly local interactions. In other words, we neglect the momentum dependence of the interaction potential, $V({\bm p}) = V_0<0$. Then the only non-vanishing component of the superconducting gap is the singlet component, $\Delta_{0s} = \Delta$, and it becomes momentum-independent. Since by a Zeeman splitting the singlet component of the gap function will be the most affected, we can make this approximation to examine the fate of the superconducting condensate with increasing Zeeman potential.

 For a strictly local attractive interaction, the superconducting spectrum is given by,
\begin{equation}
E_{1(2)}^2({\bm k}) = \xi_{\bm k}^2  +\alpha_k^2 +\Gamma^2 + |\Delta|^2 \mp 2 \sqrt{\xi_{\bm k}^2\alpha_k^2 + \Gamma^2( \xi_{\bm k}^2 +  |\Delta|^2)}, \label{E12}
\end{equation}
where $\alpha_k =\alpha k$.
Solving the kinetic equations for ${\cal F}_{++}$ and ${\cal F}_{--}$ and using Eq (\ref{del0s}) we obtain the  gap equation for the strictly local attractive interaction,
\begin{equation}
  k_B T\sum_n \frac{-V_0}{2}\sum_{\bm q}\left[\frac{1}{\omega_n^2 +E_1^2} + \frac{1}{\omega_n^2 +E_2^2}  -\frac{\Gamma^2}{\sqrt{\xi_{\bm q}^2\alpha_q^2 + \Gamma^2(\xi_{\bm q}^2 +|\Delta|^2)}} \left(\frac{1}{\omega_n^2 +E_1^2} - \frac{1}{\omega_n^2 +E_2^2} \right) \right]=1.
\end{equation}
Taking the zero temperature limit and performing the summation over the frequencies we obtain
\begin{equation}
1= \frac{-V_0}{2}\sum_{\bm q}\left[\frac{1}{2E_1({\bm q})} +  \frac{1}{2E_2({\bm q})}- \frac{\Gamma^2}{\sqrt{\xi_{\bm q}^2\alpha_q^2 + \Gamma^2(\xi_{\bm q}^2 +|\Delta|^2)}} \left(\frac{1}{2E_1({\bm q})} -  \frac{1}{2E_2({\bm q})} \right) \right]. \label{GapEq}
\end{equation}

~
\section{Anomalous correlation functions and gap functions in the c--operator representation}

 To obtain a deeper understanding of the singlet--triplet mixing in superconductors with spin--orbit coupling \cite{Gorkov}, it is useful to determine the expressions of the anomalous  correlation functions and of the gap functions in terms of the original electron operators.  We first express the $c$--operators in terms of $a$--operators, $c_{{\bm p}\uparrow}=(a_{{\bm p}+} + a_{{\bm p}-})/\sqrt{2}$ and $c_{{\bm p}\downarrow}=-i e^{i\theta_{\bm p}} (a_{{\bm p}+} - a_{{\bm p}-})/\sqrt{2}$, and we obtain for the singlet and triplet anomalous correlation functions the expressions
\begin{eqnarray}
\langle T_{\tau} c_{-{\bm p} \downarrow}(\tau) c_{{\bm p} \uparrow}(\tau^{\prime}) \rangle &=& \frac{i}{2} e^{i\theta_{\bm p}} \left[ {\cal F}_{++}(-{\bm p}, \tau-\tau^{\prime}) + {\cal F}_{--}(-{\bm p}, \tau-\tau^{\prime}) + {\cal F}_{+-}(-{\bm p}, \tau-\tau^{\prime}) + {\cal F}_{-+}(-{\bm p}, \tau-\tau^{\prime})\right], \\
\langle T_{\tau} c_{-{\bm p} \sigma}(\tau) c_{{\bm p} \sigma}(\tau^{\prime}) \rangle &=& \frac{1}{2} e^{(1-\sigma)i\theta_{\bm p}} \left[ {\cal F}_{++}(-{\bm p}, \tau-\tau^{\prime}) - {\cal F}_{--}(-{\bm p}, \tau-\tau^{\prime}) +\sigma {\cal F}_{+-}(-{\bm p}, \tau-\tau^{\prime}) -\sigma {\cal F}_{-+}(-{\bm p}, \tau-\tau^{\prime})\right],  \nonumber
\end{eqnarray}
where the ${\cal F}$ anomalous functions are given by equations (14) and (15). In the limit of local interactions we can determine the explicit dependence of the ${\cal F}$ functions on the parameters of the  model using Eq. (21), and we have $({\cal F}_{\lambda\lambda^\prime} +{\cal F}_{-\lambda -\lambda^\prime}) \propto \Delta_{0s}({\bf p})$ and $({\cal F}_{\lambda\lambda^\prime} - {\cal F}_{-\lambda -\lambda^\prime}) \propto  i \alpha p e^{-i\theta_{\bm p}} \Delta_{0s}({\bf p})$. Consequently, in the $c$--operator representation both the singlet and the triplet components of the anomalous correlation function are proportional to the $s$--wave gap, $\langle T_{\tau} c_{-{\bm p} \downarrow}(\tau) c_{{\bm p} \uparrow}(\tau^{\prime}) \rangle \propto  \Delta_{0s}({\bf p})$ and $\langle T_{\tau} c_{-{\bm p} \sigma}(\tau) c_{{\bm p} \sigma}(\tau^{\prime}) \rangle \propto  i \alpha\Delta_{0s}({\bf p})(p_x - i \sigma p_y)$, respectively.

We emphasize that, in the limit of local pairing interaction, the anomalous correlation function in the $c$--operator representation  has both singlet and triplet components, but the corresponding gap function is purely $s$--wave.  To show this property explicitly, we can re-derive the gap equations in the $c$--operator representation and, instead of Eq. (20),  we obtain
\begin{equation}
(-\partial_{\tau} -\xi_{\bm p} -\sigma\Gamma)G_{\sigma\sigma^\prime}({\bm p}, \tau-\tau^\prime) -\sigma \alpha p  G_{- \sigma\sigma^\prime}({\bm p}, \tau-\tau^\prime) +\sum_{\sigma_1} \Delta_{\sigma_1 \sigma}({\bm p})\langle T_{\tau} c_{{\bm p}\sigma^\prime}^{\dagger} c_{-{\bm p}\sigma_1}^{\dagger} \rangle = \delta_{\sigma\sigma^\prime}\delta(\tau -\tau^\prime),
\end{equation}
where $G_{\sigma\sigma^\prime}({\bm p}, \tau-\tau^\prime)=-\langle T_{\tau} c_{{\bm p} \sigma}(\tau) c_{{\bm p} \sigma^\prime}^\dagger(\tau^{\prime}) \rangle$ is the normal Green function and the gap functions are defined as
\begin{equation}
\Delta_{\sigma\sigma^\prime}({\bm p}) = -\sum_{\bm q} V({\bm p}-{\bm q}) \langle  c_{-{\bm p} \sigma}(\tau) c_{{\bm p} \sigma^\prime}(\tau) \rangle.
\end{equation}
The equal time anomalous correlation $\langle  c_{-{\bm p} \sigma}(\tau) c_{{\bm p} \sigma^\prime}(\tau) \rangle$ can be expressed in terms of ${\cal F}_{\lambda\lambda^\prime}(0)$. Explicitly, we have
\begin{eqnarray}
\Delta_{\downarrow \uparrow}({\bm p}) &=& -\sum_{\bm q}\frac{V({\bm p}-{\bm q})+ V({\bm p}+{\bm q})}{2}\frac{i}{2}e^{i\theta_{\bm q}}\left[{\cal F}_{++}+{\cal F}_{--}\right] -\sum_{\bm q}\frac{V({\bm p}-{\bm q})- V({\bm p}+{\bm q})}{2}\frac{i}{2}e^{i\theta_{\bm q}}\left[{\cal F}_{+-}+{\cal F}_{-+}\right], \nonumber \\
 \Delta_{\uparrow \uparrow}({\bm p}) &=& -\sum_{\bm q}\frac{V({\bm p}-{\bm q})- V({\bm p}+{\bm q})}{2}\frac{1}{2}\left[{\cal F}_{++}-{\cal F}_{--}\right] -\sum_{\bm q}\frac{V({\bm p}-{\bm q})+ V({\bm p}+{\bm q})}{2}\frac{1}{2}\left[{\cal F}_{+-}-{\cal F}_{-+}\right].\nonumber
 \end{eqnarray}
In the limit of strictly local interactions $V({\bm p}-{\bm q})= V({\bm p}+{\bm q})=V_0$ and $({\cal F}_{+-}-{\cal F}_{-+}) \propto q_x-i q_y$ and, consequently, the triplet component vanishes, $\Delta_{\uparrow\uparrow}=\Delta_{\downarrow\downarrow}=0$. On the other hand, the expression for the singlet component of the gap becomes identical with the right hand side of Eq. (22),  hence we have $\Delta_{\downarrow \uparrow}({\bm p}) = \Delta_{0s}({\bm p}) = \Delta$. We conclude that in a superconductor with spin--orbit coupling and on-site pairing interactions the anomalous correlation function is characterized by a mixture of singlet and triplet components, yet the gap function has purely $s$--wave symmetry. A $p$--wave component of the gap can develop only in the presence of non--local pairing interactions. As a consequence, in a system with strictly local pairing interaction, if the singlet anomalous correlation vanishes, the superconducting gap as well as  all the other components of the  anomalous correlation function will vanish.

\section{Analysis of the gap equation}
We first analyze the gap equation, Eq.~(\ref{GapEq}), in some special cases for which the solutions are well known.
This will serve as a test for the validity of our analytical calculations.
By putting $\Gamma=0$ and $\alpha=0$, which corresponds to the standard BCS case of a local attractive interaction with no spin-orbit coupling and Zeeman splitting, we find $E_1 = E_2 = \sqrt{\xi_q^2 + |\Delta|^2}$.
In this case, from Eq.~(\ref{GapEq}) we recover the standard BCS gap equation,
\begin{equation}
1= \frac{-V_0}{2}\sum_{\bm q}\frac{1}{\sqrt{\xi_q^2 + |\Delta|^2}},
\label{CaseI}
\end{equation}
where the summation over ${\bm q}$ should be performed over states satisfying $|\xi_{\bm q}| < \omega_D$, with $\omega_D$ some cut-off Debye energy scale. As is well known \cite{Schrieffer-Book}, since the integral on the right hand side (R.H.S.) diverges in the limit $|\Delta| \rightarrow 0$, in this case a non-zero solution for $\Delta$ exists for any $V_0<0$.

Next we take the system with $\Gamma=0$, $\alpha\neq0$. In this case $E_{1(2)} =\sqrt{ (\xi_{\bm q}\mp\alpha_q)^2+|\Delta|^2}$. The gap equation now becomes,
\begin{equation}
1= \frac{-V_0}{4}\sum_{\bm q}\left[\frac{1}{\sqrt{ (\xi_{\bm q}-\alpha_q)^2+|\Delta|^2}} + \frac{1}{\sqrt{ (\xi_{\bm q}+\alpha_q)^2+|\Delta|^2}}\right]
\label{CaseII}
\end{equation}
As in the previous case, the integrals on the R.H.S. of Eq.~(\ref{CaseII}) diverge when $|\Delta| \rightarrow 0$, hence a non-vanishing solution for $\Delta$ exists for any $V_0 < 0$.

To establish the familiar result that $s$-wave superconductivity is destroyed by Zeeman splitting (in the absence of spin-orbit coupling), we consider the special case,  $\Gamma\neq0$, and $\alpha=0$. In this case $E_{1(2)} =|\sqrt{\xi_{\bm q}^2+|\Delta|^2} \mp \Gamma|$, and the gap equation becomes
\begin{equation}
1= \frac{-V_0}{2}\sum_{\bm q}\frac{1}{\sqrt{\xi_q^2 + |\Delta|^2}},
\label{CaseIII}
\end{equation}
where the summation over ${\bm q}$ is done over states satisfying $|\xi_{\bm q}| < \omega_D$ and $\sqrt{\xi_{\bm q}^2+|\Delta|^2} > \Gamma$.
The second constraint results from the cancelations of two terms from Eq.~(\ref{GapEq}) that diverge in the limit $|\Delta| \rightarrow 0$. Since the integral on the R.H.S. of the gap equation no longer diverges, a non-zero solution for $\Delta$ exists only for
$|V_0|$ larger that a critical value. This implies that for a given strength of the attractive potential $|V_0|$, no non-zero solution for $\Delta$ can be found above a critical value of the Zeeman potential $\Gamma$.

Finally we consider the most general case of a non-zero Zeeman potential as well as a non-zero spin-orbit coupling, $\Gamma\neq0$, $\alpha\neq0$. The gap equation is then given by Eq.~(\ref{GapEq}).
The exact cancelation of the divergent terms that characterizes the $\alpha=0$ case does no longer hold and  the R.H.S. of Eq.~(\ref{GapEq}) becomes  arbitrarily large in the limit $|\Delta| \rightarrow 0$. Consequently, a non-vanishing solution for $\Delta$ exists for \emph{any} negative value of $V_0$.  This implies that $\Delta$ does not vanish for any value of $\Gamma$, or, in other words, the pair potential cannot be completely destroyed by a Zeeman splitting  in the presence of a non-zero spin-orbit
coupling. This is in agreement with a similar result derived earlier in a different context \cite{Frigeri}. Nonetheless, at large values of $\Gamma$ the superconducting pair potential decreases exponentially with the strength of the Zeeman splitting as we show in the next section.

\section{Numerical solution and quantum phase transitions}

Next, we determine the general solution of the gap equation by solving Eq. (\ref{GapEq}) numerically. We address two distinct cases: 1) High carrier concentration regime, when the chemical potential $\mu$ (i.e., the Fermi energy in our zero temperature limit) represents the largest energy scale in the problem, $\mu\gg \omega_D, \Gamma, \alpha k_F, \Delta(0)$, and 2) Low carrier concentration, when $\omega_D> \mu, \Gamma, \alpha k_F, \Delta(0)$.  Here $\omega_D$ is the analog of the Debye frequency (i.e., the characteristic energy cut-off for the intrinsic pairing interaction), $ \alpha k_F$ is the strength of the spin-orbit interaction at the Fermi wave vector, and  $\Delta(0)$ is the value of the superconducting gap at zero Zeeman splitting. The zero field gap is a measure of the pairing interaction strength, and in fact one can use $V_0$, instead of $\Delta(0)$, as an independent parameter. The Debye frequency acts as a cutoff in Eq. (\ref{GapEq}), i.e.,  the summation  over ${\bm q}$ of a function $f({\bm q}, E_i({\bm q}))$ is restricted to the values of the wave vector satisfying $E_i({\bm q})<\omega_D$.

\begin{figure}[tbp]
\begin{center}
\includegraphics[width=0.45\textwidth]{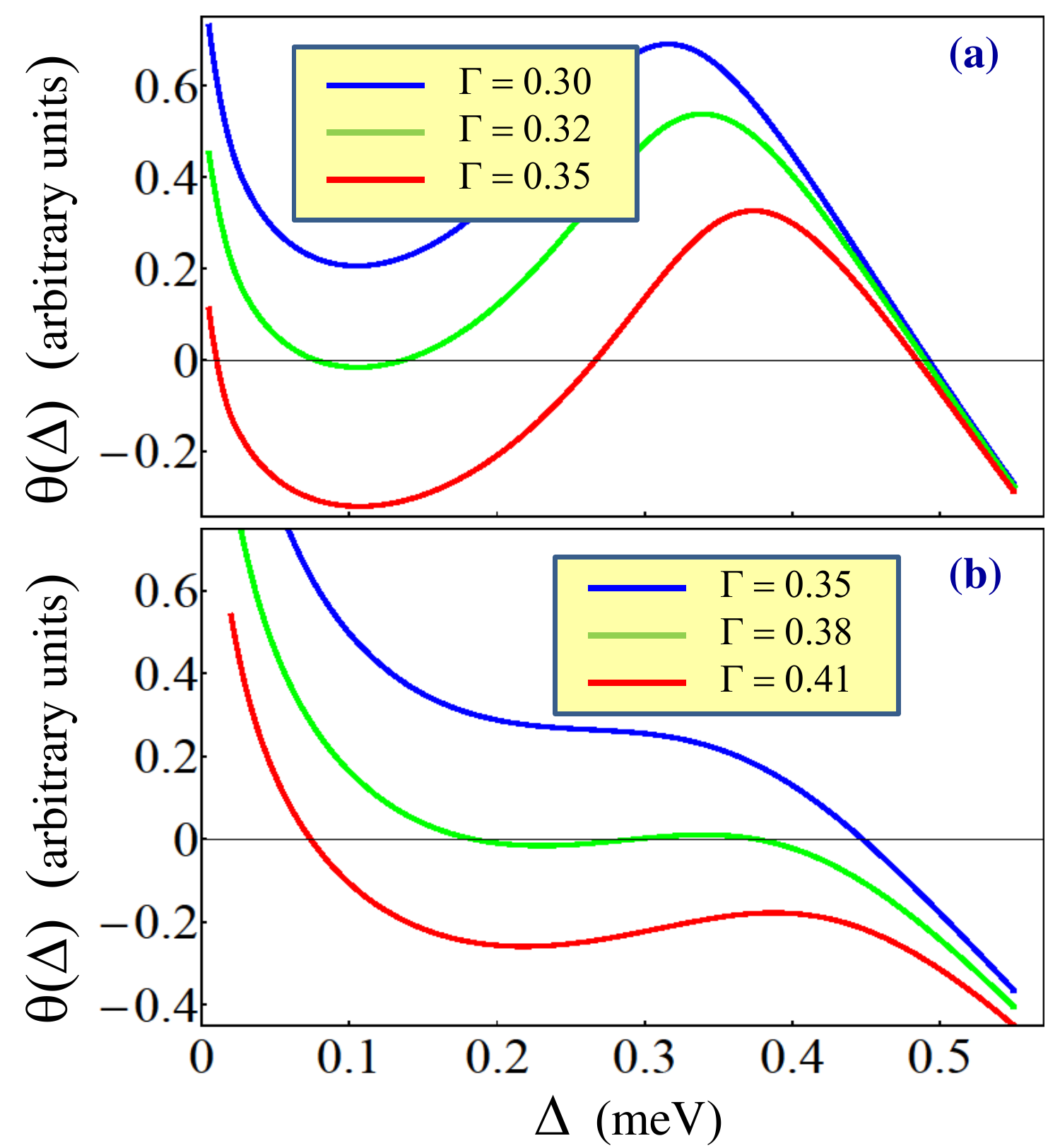}
\end{center}
\caption{(Color online) The behavior of the function $\theta(\Delta)$ (Eq.~(\ref{theta})) with $\Delta$. The values of $\Delta$ at which $\theta(\Delta)=0$ are the solutions of the self-consistent gap equation. (a): $\theta(\Delta)$ with $\Delta$ for various values of the Zeeman splitting $\Gamma$ for a small value of the spin-orbit coupling constant $\alpha$ ($2\alpha k_F = 0.2 meV$) and large $\mu = 0.5 eV$. Initially for smaller values of the Zeeman splitting there is just one solution for $\Delta$. With increasing values of $\Gamma$, there are three solutions for $\Delta$, one of which (the one in the middle) is unstable. The lower (higher) solution for $\Delta$ is the high (low) field solution which co-exist in a region of co-existence. With $\Gamma$ going up, the high field solution becomes smaller but is never zero. Consequently, there is no true first order transition in the presence of a small $\alpha$, even though there is a first-order-like precipitous drop of $\Delta$ at some value of $\Gamma$, which is a remnant of the first order transition for $\alpha=0$.  (b): $\theta(\Delta)$ with $\Delta$ for increasing values of $\Gamma$ with a fixed large $\mu=0.5 eV$ and \emph{a larger spin orbit coupling} $2\alpha k_F =0.4 meV$. $\theta(\Delta)$ is now monotonically decreasing and the resultant unique solution for $\Delta$ decreases for increasing values of $\Gamma$. The rate of decrease of this solution with $\Gamma$ is much slower and continuous. As with the case of smaller $\alpha$, there is no quantum phase transition.}
\label{Fig1}
\end{figure}

To obtain the self-consistent numerical solution for the gap, we define the function
\begin{equation}
 \theta(\Delta) = \frac{-V_0}{2}\sum_{\bm q}\left[\frac{1}{2E_1({\bm q})} +  \frac{1}{2E_2({\bm q})}- \frac{\Gamma^2}{\sqrt{\xi_{\bm q}^2\alpha_q^2 + \Gamma^2(\xi_{\bm q}^2 +|\Delta|^2)}} \left(\frac{1}{2E_1({\bm q})} -  \frac{1}{2E_2({\bm q})} \right) \right]-1. \label{theta}
\end{equation}
With this notation, Eq.  (\ref{GapEq}) becomes $\theta(\Delta) = 0$. This equation is characterized by two qualitatively different regimes that are controlled by the relative strength of the spin-orbit interaction and the zero-field gap. If $2\alpha k_F \gtrsim \Delta(0)$, $\theta(\Delta)$ is a monotonically decreasing function that starts from large positive values at $\Delta\rightarrow 0$ and  Eq. (\ref{theta}) has always one non-vanishing solution. By contrast, when  $2\alpha k_F < \Delta(0)$, the function $\theta(\Delta)$  becomes non-monotonic for certain values of the Zeeman field $\Gamma$, which means that Eq.  (\ref{theta}) can have multiple non vanishing solutions for a given set of parameters. To illustrate this situation, we show in Fig. \ref{Fig1} the function $\theta(\Delta)$ for a system with large carrier concentration ($\mu = 0.5 eV$) and extremely low spin-orbit interaction ($\alpha =0.5 meV\cdot$\AA in panel (a) and  $\alpha =1 meV\cdot$\AA in panel (b), i.e., $2\alpha k_F = 0.2 meV$ and  $2\alpha k_F = 0.4 meV$, respectively).  The Debye frequency is  $\omega_D = 25 meV$ and the zero field gap is $\Delta(0) = 0.5 meV$.  At low Zeeman splitting,   $\theta(\Delta)$ vanishes at a single point $\Delta\approx 0.5 meV$, but increasing $\Gamma$ leads to a local minimum in  $\theta(\Delta)$ that goes to zero for $\Gamma \approx 0.318 meV$. Further increasing the Zeeman splitting leads to three non-vanishing solutions $\Delta_1(\Gamma) > \Delta_2(\Gamma) > \Delta_3(\Gamma)$ (see Fig. \ref{Fig1}), where  $\Delta_1(\Gamma)$ and  $\Delta_3(\Gamma)$ are the ``low field'' and ``high field'' solutions, respectively, and $\Delta_2(\Gamma)$ is an unstable solution. The ``low-'' and ``high-field'' solutions coexist in some range of Zeeman field strengths suggesting that the system undergoes a precipitous drop in $\Delta$ akin to a field-tuned first order phase transition. The coexistence region shrinks as the strength of the spin-orbit coupling increases (see Fig. \ref{Fig1}b) and vanishes at a value $\alpha_c\approx 1.1 meV\cdot$\AA. We note that in real systems such as non-centrosymmetric superconductors the strength of the spin-orbit coupling is usually larger than this critical value and, consequently, the first-order-like precipitous drop in $\Delta$ may not be observable. In cold fermion systems the spin-orbit coupling constant can be used as a tuning parameter to interpolate between these two behaviors.

\begin{figure}[tbp]
\begin{center}
\includegraphics[width=0.45\textwidth]{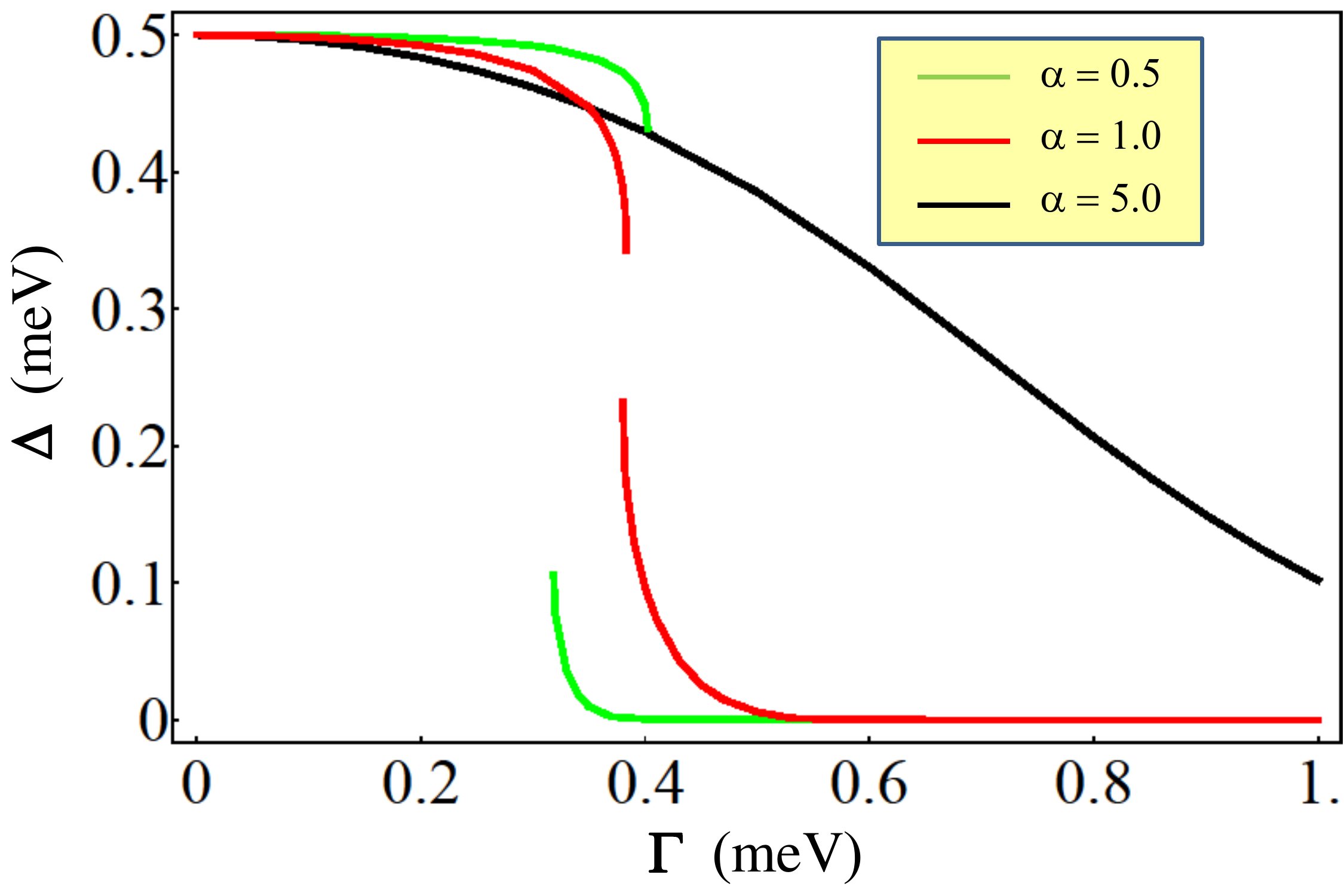}
\end{center}
\caption{(Color online) The solution of the gap equation (see Fig.~\ref{Fig1}) plotted against the Zeeman splitting $\Gamma$ for various values of the spin-orbit coupling constant $\alpha$ (in $meV\cdot$\AA). For small values of $\alpha$, the superconducting gap falls discontinuously with $\Gamma$ but it is never strictly zero in the presence of the SO coupling. Consequently, there is only a first-order-like crossover which is a remnant of the true first order phase transition with $\Gamma$ for $\alpha=0$. For larger values of $\alpha$ (black curve), the decay of the superconducting pair potential with the Zeeman splitting is much slower and continuous. It falls exponentially (but is never strictly zero) only for higher fields $\Gamma > 2\alpha k_F$.}
\label{Fig2}
\end{figure}

The dependence of the solution of the gap equations on $\Gamma$ is shown in Fig. \ref{Fig2}. The coexistence region can be easily seen for $\alpha=0.5 meV\cdot$\AA (green line in Fig \ref{Fig2}), corresponding to the (stable) solutions of the equation $\theta(\Delta)=0$ for the $\theta$ function shown in Fig. \ref{Fig1}a. Note the exponential decay of the ``high field'' solution $\Delta_3$ with increasing $\Gamma$. Practically the superconducting gap is negligible ($\Delta < 1 \mu eV$) for $\Gamma > 0.41 meV$. The coexistence region shrinks as we approach the critical spin-orbit coupling (red line in Fig. \ref{Fig2} and Fig. \ref{Fig1}b), then,  for $\alpha>\alpha_c$ the gap equation has a continuous solution $\Delta(\Gamma)$ that decreases monotonically with the Zeeman field (black line in Fig \ref{Fig2}). Note that at high fields,  $\Gamma > 2\alpha k_F$, the gap decreases exponentially. However, the energy scale for the spin-orbit coupling, $2\alpha k_F$, can be significant in realistic systems (tens of meV) and the high field regime may not be attainable, i.e., the gap will not vanish for any realistic value of the Zeeman field.

The existence of a first-order-like drop in $\Delta$ that ends at a critical value of the SO coupling $\alpha_c$ is generic, i.e., this feature is present at any value of the carrier density. However, to realize a topologically nontrivial non-Abelian regime, it is necessary to satisfy the condition $\Gamma^2 >\mu^2 +\Delta^2$. Consequently, we study the solutions of the gap equation in the low density regime, when the chemical potential, the Zeeman field, the spin-orbit interaction, and the superconducting order parameter are comparable. \emph{In particular we address the following question: is it possible to realize the condition for the existence of a topologically nontrivial non-Abelian phase while maintaining a reasonable superconducting gap?} Before presenting the results, we note that in the low-density regime the zero-field gap has a strong dependence on the chemical potential. More precisely, for a given set of parameters $V_0$, $\omega_D$, and $\alpha$, the zero-field gap $\Delta(0)$ decreases with $\mu$. In our calculations we fix $V_0$ at a value that corresponds to $\Delta(0)=0.4 meV$ at $\mu=2 meV$ and, at lower carrier densities (i.e., lower values of $\mu$), we calculate the zero-field gap using the gap equation. Also, we note that, as we vary the Zeeman splitting $\Gamma$,  the chemical potential of a system with fixed carrier density $n$ remains constant as long as the high-energy band $E_2$ has non-zero occupation. For higher values of $\Gamma$, i.e., when the bands split, we determine the chemical potential $\mu=\mu(n, \Gamma)$ corresponding to the fixed carrier density. The values of $\mu$ provided below represent zero field values.

\begin{figure}[tbp]
\begin{center}
\includegraphics[width=0.45\textwidth]{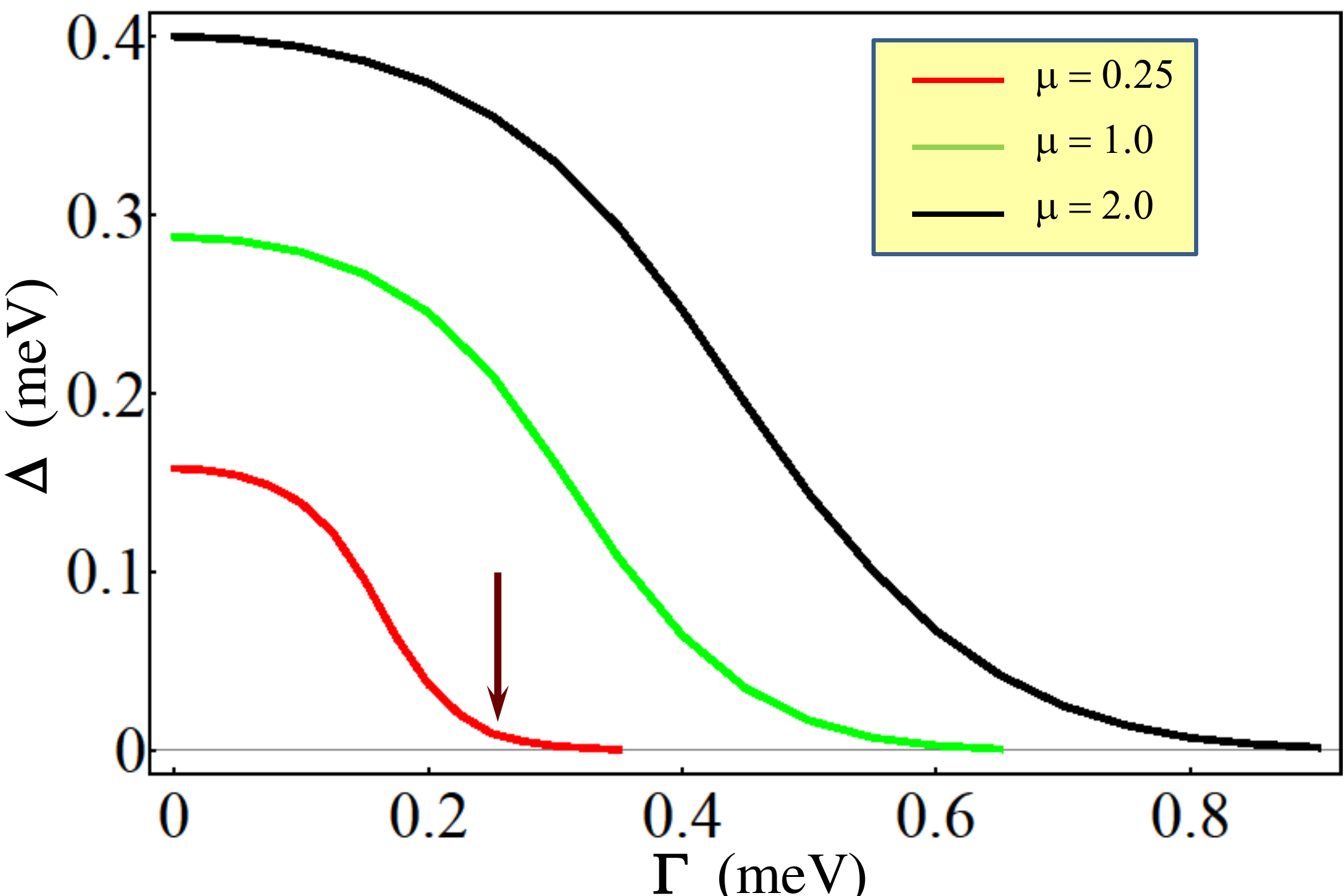}
\end{center}
\caption{(Color online) The solution of the gap equation plotted against the Zeeman splitting $\Gamma$ for three different values of $\mu$ for a fixed value of the spin-orbit coupling constant $\alpha=0.1 eV\cdot$\AA. The value of $\alpha$ is large enough so that $\Delta$ remains continuous with $\Gamma$. For larger values of $\mu$ (black and green curves), $\Delta$ becomes inappreciably small for $\Gamma \geq \mu$. For the red curve, however, $\Delta$ is appreciable ($~ 0.02$ meV) for $\Gamma \geq \mu = 0.25$ meV. Since to the right of this $\Gamma$ it is possible to satisfy
 $\Gamma^2 > \mu^2 + \Delta^2$, the system is in a topologically non-trivial phase in this region. Therefore, somewhere above $\Gamma = 0.25$ meV (shown with an arrow), there is a topological quantum phase transition from a regular superconducting phase (to the left of the arrow) to a topologically non-trivial non-Abelian phase (to the right of the arrow).}
\label{Fig3}
\end{figure}

\begin{figure}[tbp]
\begin{center}
\includegraphics[width=0.45\textwidth]{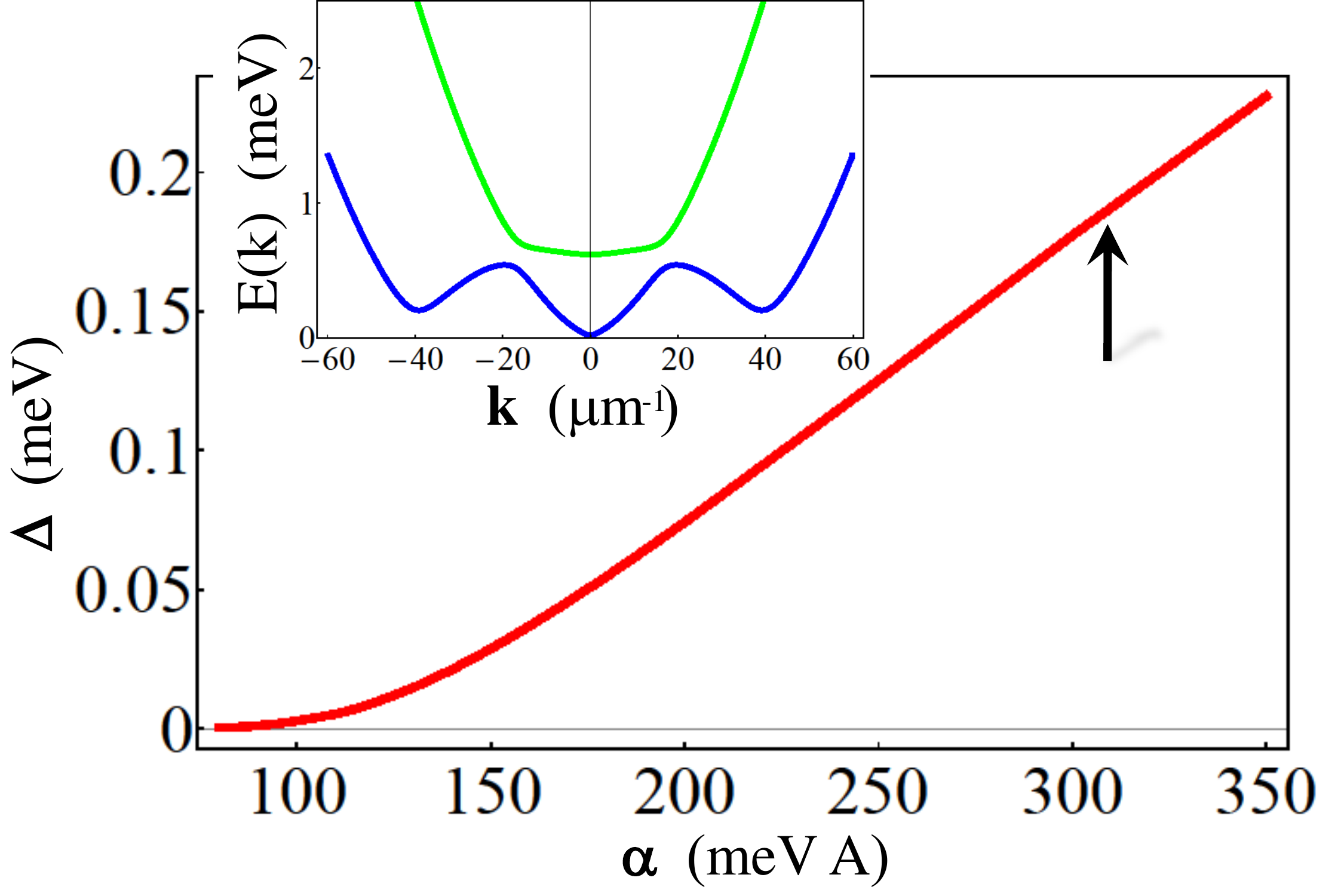}
\end{center}
\caption{(Color online) Dependence of the superconducting pair potential $\Delta$ in the topological phase on the strength of Rashba spin-orbit coupling $\alpha$. The system  is characterized by a chemical potential at zero magnetic field $\mu_0 = 0.25 meV$ (see the red curve in Fig.~\ref{Fig3} for the dependence of $\Delta$ on $\Gamma$). The Zeeman splitting is fixed at $\Gamma=0.3 meV$. With these parameters the system is in a non-Abelian phase for $\alpha=0.1 eV\cdot$\AA. The superconducting pair potential increases monotonically with $\alpha$. For $\alpha\geq 303 meV\cdot$\AA (black arrow)
the condition $\Gamma^2>\mu^2+\Delta^2$ is no longer satisfied and the system undergoes another topological quantum phase transition to an ordinary superconducting phase (at large $\alpha$). The inset shows the two excitation energies, Eq.~(\ref{E12}), of which the smaller one (blue curve) vanishes at $k=0$ at this $\alpha$-tuned TQPT.}
\label{Fig4}
\end{figure}

 Shown in Fig. \ref{Fig3} is the dependence of the solution of the gap equation on the Zeeman splitting for three different values of the chemical potential (i.e., three carrier densities) $\mu=2.0 meV, 1 meV, 0.25 meV$.  The Debye frequency is taken as $\omega_D=25 meV$ and the Rashba coupling is $\alpha = 0.1 eV\cdot$\AA,  i.e., the system is characterized by a strong spin-orbit coupling. For these parameters the system is above the critical value of $\alpha$ for the discontinuous fall of $\Delta$ and hence $\Delta$ is now a continuous function of $\Gamma$. Before analyzing the plots in Fig.~\ref{Fig3}, let us remind ourselves that in order to satisfy the conditions for the non-Abelian $s$-wave phase ($\Gamma^2 >\mu^2 +\Delta^2$), we need an appreciable $\Delta$ when $\Gamma$ has crossed $\sim \mu$. Coming back to Fig.~\ref{Fig3} we note that, similar to Fig.~\ref{Fig2} (black curve), $\Delta$ falls with increasing values of $\Gamma$. For the black and the green curves (higher $\mu$) $\Delta$ becomes inappreciably small (though it is never zero) by the time $\Gamma$ becomes $\sim \mu$.
 However, for the red curve ($\mu=0.25 meV$) there is a residual superconducting pair potential $\Delta\approx 0.02meV$ for $\Gamma\geq \mu$, i.e., the system is in a topologically nontrivial non-Abelian phase. Moreover, as shown in Fig.~\ref{Fig4}, the magnitude of this residual $s$-wave pair potential increases with $\alpha$, and thus can be increased by increasing the value of the spin-orbit coupling. Therefore, for these parameter values, there is topological quantum phase transition (TQPT) when $\Gamma$ crosses the critical value $\Gamma_c=\sqrt{\mu^2+\Delta^2}$ (shown with an arrow in Fig.~\ref{Fig3}). The TQPT separates a regular (non-topological) superconducting phase ($\Gamma < \Gamma_c$) from a topological non-Abelian superconducting phase ($\Gamma > \Gamma_c$). From our self-consistent mean field theory we find this TQPT to be continuous, that is, there is no change in $\Delta$ at the critical value of the Zeeman splitting.

\section{TQPT in the proximity induced case}
An alternative and perhaps more robust way to create a topologically non-trivial non-Abelian superconductor is to induce
a superconducting pair potential in a spin-orbit coupled semiconductor by proximity
effect \cite{sau1,Ann,Long-PRB}.
Ideally, for the proximity effect induced
superconductivity the pairing interaction resides in a
parent $s$-wave superconductor such as Al or Nb while the
quasiparticles of interest are confined to a two- or one-dimensional
semiconductor layer
on the surface of the superconductor.
 The proximity effect has been shown to create
a topological superconductor similar to the ones discussed
above on the surface state of a topological insulator
 \cite{fu_prl'08,robustness,tudor_TI} and also in a 2D semiconductor layer
 \cite{Long-PRB}.

Physically, the proximity effect arises from  multiple
Andreev reflections of electrons in a semiconductor that is
connected to a superconductor by tunneling. For most
realistic cases, there is no pairing interaction
 in the semiconductor. Thus, strictly speaking the superconducting
 pair potential vanishes in the
semiconductor and at first glance it appears that there is no
superconductivity induced in the semiconductor.
However, the superconducting order parameter defined by
 $\langle\psi_\sigma^\dagger(\bm r)\psi_{\sigma'}^\dagger(\bm r')\rangle$
 is found to remain non-zero in the semiconductor layer. Furthermore,
the multiple Andreev reflections open a gap in the
spectrum of quasiparticles that are localized in the semiconductor
layer. The spectra of such quasiparticles can be shown to be
identical to quasiparticles with an \emph{effective} pairing potential in the
semiconductor layer \cite{robustness}. Therefore from the point of the
quasiparticle spectrum, which is the only property that is relevant
to the definition of a topological superconductor, the proximity to a
superconductor induces a superconducting quasiparticle gap in the
semiconductor.

\begin{figure}[tbp]
\begin{center}
\includegraphics[width=0.45\textwidth]{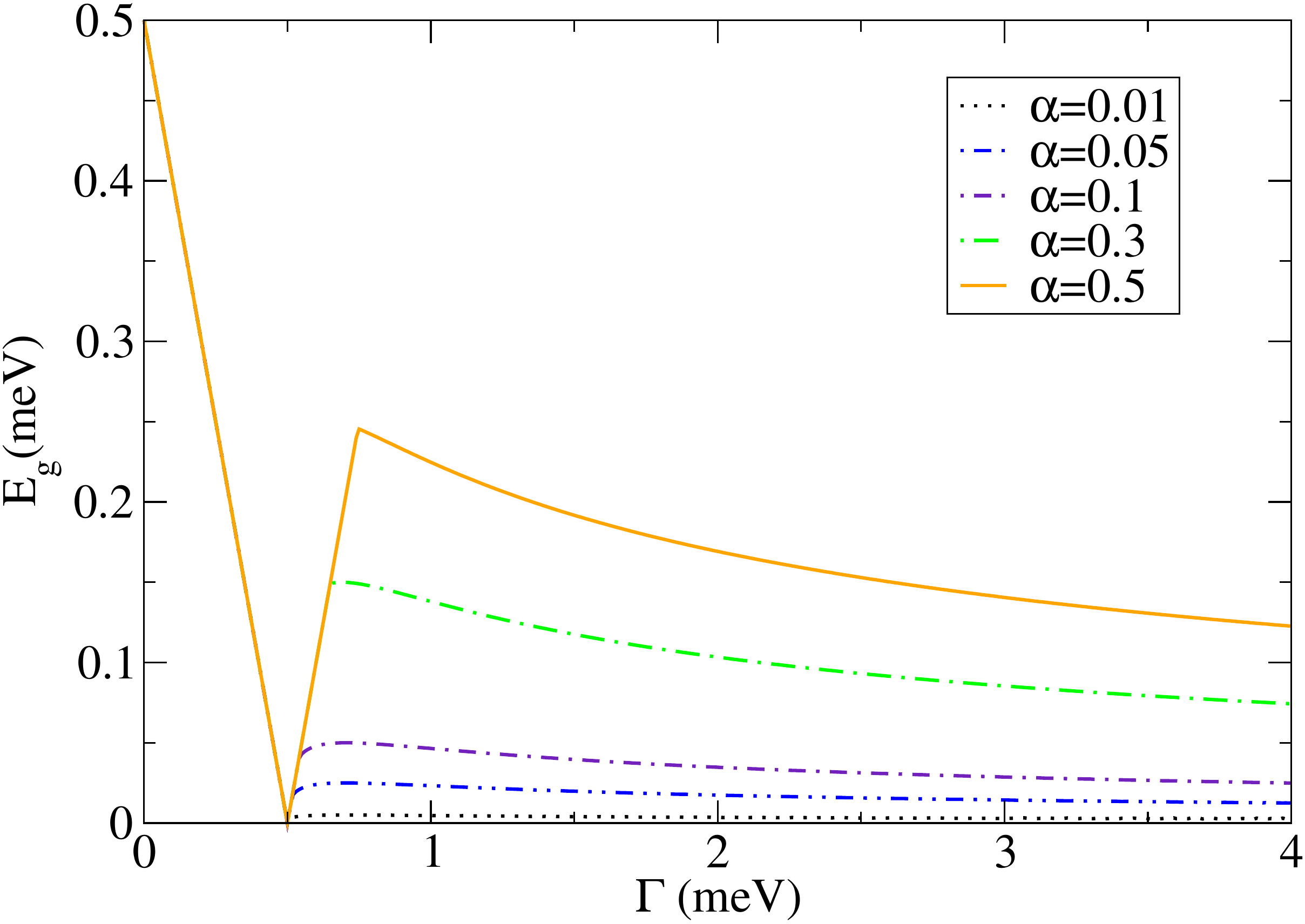}
\end{center}
\caption{(Color online)
Quasiparticle gap $E_g$ versus Zeeman coupling $\Gamma$ for various
 values of spin-orbit interaction $\alpha$. The strength of the
 spin-orbit coupling in the inset is such that $\alpha=0.3$
 corresponds to $0.1$ eV-\AA. The proximity induced pair potential
and chemical potential are taken to be $\Delta_{eff}=0.5 meV$ and $\mu=0.0$.
 The quasiparticle gap vanishes at the critical value $\Gamma_{c}=\sqrt{\Delta_{eff}^2+\mu^2}$.
  Above the critical point spin-orbit coupling opens a quasiparticle
gap that is proportional to $\alpha$ in the small $\alpha$ limit.
}
\label{Fig5}
\end{figure}

The proximity effect can be induced by even weak tunneling between
the semiconductor and the superconductor. Therefore, the quasiparticle
spectrum in the semiconductor does not affect the pairing potential
in the superconductor significantly. \emph{Specifically, for the
proximity induced superconductivity case, the self-consistency effects
that were important in the discussions in the previous sections
become insignificant.} Furthermore, if the Zeeman potential is
also induced by proximity effect from a magnetic insulator on the
other surface of the semiconductor, there is no direct tunneling
between the superconductor and the magnetic insulator and therefore
no suppression of the order parameter in the
superconductor \cite{Long-PRB}.
Thus, in contrast to the discussions in the previous sections,
where the Zeeman potential induced topological phase transition
 was accompanied by significant changes in the pair
 potential $\Delta$, the pair potential $\Delta$ in the proximity
induced case remains unaffected by the Zeeman splitting.

The TQPT in both cases (the proximity induced case and the case when the pair potential is due to an intrinsic pairing interaction) can be characterized by the closing of the
superconducting quasiparticle gap (shown in Fig.~\ref{Fig5})
 as the Zeeman potential
is raised from $\Gamma=0$ past the critical value $\Gamma_c=\sqrt{\Delta^2+\mu^2}$.
In the proximity induced case, $\Delta$ is the proximity induced effective pair
potential and $\mu$ is the fermi energy in the semiconductor.
The quasiparticle gap $E_g(\bm{k})$ (minimum of $E_{1(2)} (\bm{k})$ in Eq.~(\ref{E12})) closes at $k=0$ exactly when $\Gamma$ passes through $\Gamma_c$ (Fig.~\ref{Fig5}), indicating the existence of a QPT even though the superconducting pair potential $\Delta$ remains perfectly continuous.
The quasiparticle gap for $\Gamma > \Gamma_c$
shows a linear dependence on the spin-orbit coupling strength $\alpha$
at small $\alpha$ \cite{Long-PRB}.
Here it is appropriate to mention a caveat for the case where the
Zeeman potential is not proximity induced and instead
induced by a magnetic field \cite{Roman,Oreg,Alicea-Tunable}.
In this case, the Zeeman potential also
suppresses the superconducting pair potential in the parent $s$-wave
superconductor. However, this effect can be small provided the
$g$-factor in the semiconductor is much larger than in the superconductor
as is often the case in 2D electron systems.

 \section{Discussion}
 Topologically non-trivial non-Abelian superconductivity can be realized in two different classes of systems. In class (a), superconductivity is proximity induced on a semiconductor (in the form of a film or a wire) which has a strong spin-orbit coupling. In class (b), superconductivity arises from intrinsic attractive pairing interaction in a system which also has a sizable spin-orbit coupling. In both cases a firm requirement for a phase transition from an ordinary superconducting phase to a topologically non-trivial superconducting phase is an externally imposed Zeeman splitting. The Zeeman splitting creates a gap in the spin orbit bands (Fig.~\ref{Fig0}). When this gap is large (Zeeman splitting is comparable to the chemical potential) so the Fermi surface lies in only the lower band, it triggers a QPT at which the system goes from a regular superconducting phase (small Zeeman splitting) to a topolgical superconducting phase (large Zeeman splitting). This value of Zeeman splitting far exceeds the value at which an ordinary $s$-wave superconductor is known to lose superconductivity due to its inability to form spin-singlet zero-momentum Cooper pairs. As we have shown above, this is where the requirement of a sizable spin-orbit coupling is important to stabilize a topological superconducting phase. Below we recapitulate and discuss the main results first for the case where the superconductivity arises from an intrinsic pairing interaction followed by the much simpler case of superconductivity arising from proximity effect.

 To discuss the various phases and the quantum phase transitions we have divided the parameter space in two distinct regimes by the relative magnitude of $\mu$ with respect to all other energy scales in the problem. In the large $\mu$ regime the underlying system always has two Fermi surfaces irrespective of the magnitude of the Zeeman splitting $\Gamma$. In the absence of the spin-orbit coupling $\alpha$, with increasing values of $\Gamma$ it becomes increasingly difficult for the system to create spin-singlet $s$-wave pair potential at zero net momentum. Ignoring the possibility of Cooper pairs with non-zero net momentum, we find that when $\Gamma$ crosses a critical value $\sim \Delta$ the system becomes non-superconducting at a first order QPT. At this transition the pair potential drops discontinuously to zero. By including a non-zero $\alpha$ we find that, surprisingly, there is always a non-zero solution of the gap equation Eq.~(\ref{GapEq}). This is because with $\alpha \neq 0$ the individual bands can no longer be viewed as carrying a single spin component. Rather, both bands now carry a minority spin amplitude along with the majority component, which allows $s$-wave superconducting pairing even for large values of $\Gamma$. If $\alpha$ is small, $2\alpha k_F \leq \Delta(0)$ where $\Delta(0)$ is the value of the order parameter for zero Zeeman splitting, there is still a precipitous drop of $\Delta$ at Zeeman splitting $\Gamma \sim \Delta(0)$ (Fig.~\ref{Fig2}). However, this is not a QPT, since, as already mentioned, $\Delta$ is never strictly zero in the presence of a non-zero $\alpha$. When $\alpha$ itself crosses a threshold value, $2\alpha k_F \geq \Delta(0)$, the first-order-like drop of $\Delta$ as a function of $\Gamma$ turns into a slower continuous decay (black curve in Fig.~\ref{Fig2}). For high values of $\Gamma \geq 2\alpha k_F$, $\Delta$ again decays exponentially with $\Gamma$. However, this high field scale, comparable to the spin-orbit strength at the Fermi surface, may not be attainable in real systems. Consequently, the $s$-wave superconducting gap may never vanish with a Zeeman coupling in the presence of strong spin-orbit coupling.

 The regime of small $\mu$ is particularly important because of the possibility of a topological phase transition. In this case, the behavior of $\Delta$ with $\Gamma$ for $\alpha=0$ (first order QPT) and small $\alpha$ (precipitous drop of $\Delta$ with $\Gamma$) remain unchanged from the case with large $\mu$. For small $\mu$, however, $\Delta (0)$ itself is small. Consequently $\alpha$ is always in the large spin-orbit coupling regime, $2\alpha k_F > \Delta(0)$. Therefore, for realistic values of $\alpha$, $\Delta$ falls only gradually with $\Gamma$ and strictly speaking is never zero (Fig.~\ref{Fig3}). Let us now recall that for a TQPT from a regular $s$-wave superconductor to a topologically non-trivial superconductor the parameters need to satisfy the condition $\Gamma > \Gamma_c = \sqrt{\Delta^2 + \mu^2} $. This implies that, for a robust non-Abelian phase, we require an appreciable $\Delta$ when $\Gamma$ becomes $\geq \mu$. From the red curve in Fig.~\ref{Fig3}, we note that for $\Gamma \sim \mu$, $\Delta$ is still appreciable, $\Delta \sim .02 meV$, and thus a stable non-Abelian phase is in principle allowed. Moreover, as shown in Fig.~\ref{Fig4}, the value of $\Delta$ for large $\Gamma$ (i.e., $\Delta$ in the non-Abelian phase) is directly related to the spin-orbit strength $\alpha$ and increases appreciably if $\alpha$ can be increased (as in a cold fermion system). Conversely, there is no non-Abelian phase ($\Delta=0$) if the system has no spin-orbit coupling.

 From our self-consistent mean field theory, we find the TQPT at $\Gamma = \Gamma_c = \sqrt{\Delta^2 +\mu^2}$ to be continuous. By this we mean that the magnitude of $\Delta$ is continuous across this transition. At $\Gamma=\Gamma_c$ the underlying system shifts from having two Fermi surfaces ($\Gamma < \Gamma_c$) to just one in the lower band ($\Gamma > \Gamma_c$). As shown in Ref.~[\onlinecite{sau1}], for $\Gamma > \Gamma_c$ a defect in the superconducting order parameter (e.g., vortex, sample edge) traps a unique zero energy bound state Majorana mode. Such a non-degenerate bound state solution is absent for $\Gamma < \Gamma_c$. The emergence of the topological bound state Majorana solution for $\Gamma > \Gamma_c$ makes
the transition a topological one. The exact location of the topological transition is indicated by the quasiparticle excitation energy $E_g(\bm{k})$ (minimum of $E_{1(2)} (\bm{k})$ in Eq.~(\ref{E12})) passing through zero. This happens at $k=0$ exactly when $\Gamma$ passes through $\Gamma_c$ (Fig.~\ref{Fig5}), indicating the existence of a QPT even though the superconducting order parameter $\Delta$ remains perfectly continuous.

When $s$-wave superconductivity is proximity induced on a semiconductor, there is no self-consistent gap
equation to be satisfied in the semiconductor. Thus there is no self-consistency effects that suppress the pair potential with the Zeeman splitting as discussed above. In this case, the semiconductor simply `inherits' the superconducting pair potential and its quasiparticle spectrum is modified accordingly. For weak tunneling between the semiconductor and the superconductor layers, the quasiparticles in the semiconductor cannot significantly influence the pair potential in the host superconductor. Therefore, the self-consistency requirement as in the discussions above can be neglected. If the Zeeman potential is also induced by the proximity effect of a magnetic insulator from the opposite side of the semiconductor, there will be minimal effect of the magnetic insulator on the $s$-wave superconductor. If the Zeeman potential is induced by a parallel magnetic field, then the effect on the host superconductor will again be minimal provided the $g$-factor in the semiconductor is larger than that in the superconductor.

\section{Conclusion} To conclude, we have considered spin-orbit coupled systems with superconductivity arising from either intrinsic on-site $s$-wave pairing interactions or from the proximity effect of an adjacent superconductor. In both cases, using BdG analysis of a postulated mean field Hamiltonian with an $s$-wave pair potential $\Delta$, it has been shown \cite{sau1,Ann,Long-PRB,Parag,Sato-Fujimoto} that when an externally imposed Zeeman splitting crosses a critical value, there is a Majorana fermion mode at a vortex core. The required Zeeman splitting, $\Gamma > \Gamma_c = \sqrt{\mu^2 + \Delta^2}$, seems to far exceed the value ($\Gamma \sim \Delta$) above which an
 $s$-wave pair potential $\Delta$ is known to vanish. This gives rise to the conceptual question if the postulated pair potential in Ref.~[\onlinecite{sau1}] and all the subsequent works on this system is spurious for $\Gamma > \Gamma_c$. If true this will indicate that the  BdG result of the Majorana fermion at a vortex core for $\Gamma > \Gamma_c$, based on the postulated mean field H \cite{sau1,Ann,Long-PRB,Parag,Sato-Fujimoto}, is a spurious mathematical result with no physical connection. In this paper we have resolved this question by showing that in the presence of spin-orbit coupling the $s$-wave pair potential can never be made strictly zero by the application of a Zeeman potential. This is in agreement with a similar result derived previously in a different context \cite{Frigeri}. When the $s$-wave pair potential arises from an intrinsic local pairing interaction, our self-consistent analysis of the gap equation reveals that the decay of the pair potential with Zeeman splitting is more gradual in the presence of spin-orbit coupling, although for large enough Zeeman splitting the decay is exponential. Thus there can be a small but finite region in the parameter space (RHS of arrow, which indicates a TQPT, on Fig. 4) where a topologically non-trivial superconducting phase can be realized.
When the $s$-wave pair potential is proximity induced on a semiconductor, there is no requirement of satisfying the self-consistent gap equations. In this case, the pair potential is simply `inherited' from the adjacent superconductor. Thus in this case the topologically non-trivial phase is much more robust than the other case where it is due to intrinsic pairing interactions.

Two final comments are in order here. For long- but finite-range pairing interactions (as opposed to local interactions as in this paper) it is well known that the spin-orbit interaction mixes $s$-wave
and $p$-wave pair potentials \cite{Gorkov}. In this case it may appear that superconductivity can evade the CC limit merely because the $p$-wave part of the pair potential can survive the strong Zeeman field, even if the $s$-wave part cannot. It is, however, incorrect to ascribe the existence of the topological superconducting state at large $\Gamma$ to this effect. As we have shown in detail in Ref.~[\onlinecite{Parag}], the topological state owes its existence solely to the survival of the $s$-wave part of the pair potential. (The Pfaffian topological invariant discussed in Ref.~[\onlinecite{Parag}] is completely insensitive to the $p$-wave part.) It is precisely to isolate and eliminate the effect of the mixed $p$-wave pair potential that in this paper we confined ourselves to a strictly local pairing interaction. The existence of the topological state at high Zeeman fields is strictly due to the survival of the $s$-wave pair potential, the physics of which is discussed in this paper and also summarized in the concluding paragraph of the introduction.

But this work is not just an academic resolution of the question of the survival of an $s$-wave pair potential in the presence of a strong Zeeman field.  It also proves that all properties of the topological state when superconductivity is proximity induced continue to hold even when superconductivity is due to local $s$-wave pairing interactions. This result is directly relevant to the case of an $s$-wave Feshbach cold atom system. Note that in this case $\Delta$ cannot just be assumed in the BdG equations (as in the case of the proximity effect framework \cite{Jiang}), but has to be calculated from the gap equations as in the present paper.


\acknowledgements{S.T. acknowledges support from DARPA-MTO Grant No: FA 9550-10-1-0497. J.D.S. and S.D.S. are supported by DARPA-QuEST, JQI-NSF-PFC, and Microsoft Q. We thank the Aspen Center for Physics, where this work was conceived, for hospitality.  S.T. and J.D.S. thank S. Chakravarty, C. M. Varma, R. Shankar, and G. Murthy for discussions and for asking many critical questions during the Aspen 2010 summer workshop \emph{Low dimensional topological systems}. These questions convinced us that the issue of the Chandrasekhar-Clogston limit in this context is not as trivial as we had always thought it to be and requires a thorough analysis.}

\end{document}